\documentclass[useAMS,usenatbib,fleqn]{mn2e}
\RequirePackage[dvips]{graphicx}

\usepackage{amssymb}

\usepackage{mathptmx}

\newcommand{\nat}{Nature}
\newcommand{\apjl}{ApJ}

\newcommand{\apj}{ApJ}
\newcommand{\mnras}{MNRAS}
\newcommand{\aap}{A{\&}A}

\newcommand{\apss}{ApSS}

\newcommand{\sovast}{Soviet Astronomy}

\title[Relating $v_{\perp}$ with $\mu/\dot{\mu}$]{Relating the Kick Velocities of Young Pulsars with Magnetic Field Growth Timescales Inferred From Braking Indices}

\author[A.\ G\"{u}neyda\c{s} and  K.\ Y.\ Ek\c{s}i]{A.\ G\"{u}neyda\c{s}$^{\dagger \ast}$ and  K.\ Y.\ Ek\c{s}i$^{\dagger}$ \\
\\
\.{I}stanbul Technical University, Faculty of Science and Letters, Department of Physics, Istanbul 34469, Turkey \\
$^{\dagger}$ E-mails: guneydas@gmail.com, eksi@itu.edu.tr \\
$^{\ast}$ Present address: Sabanc\i\ University, \.{I}stanbul, Turkey
} 

\date{}

\pagerange{\pageref{firstpage}--\pageref{lastpage}} \pubyear{2012}

\begin{document}

\maketitle

\label{firstpage}

\begin{abstract}
A nascent neutron star may be exposed to fallback accretion soon after the proto-neutron star stage. 
This high accretion episode can submerge the magnetic field deep in the crust. 
The diffusion of the magnetic field back to the surface will take hundreds to millions of years 
depending on the amount of mass accreted and the consequent depth the field is buried.
Neutron stars with large kick velocities will accrete less amount of fallback material leading to 
shallower submergence of their fields and shorter time-scales for the growth of their fields. 
We obtain the relation $\tau_{\rm Ohm} \propto v^{-1}$ between the space velocity of the neutron star 
and Ohmic time-scale for the growth of the magnetic field. 
We compare this with the relation between the measured transverse velocities, $v_{\perp}$ 
and the field growth time-scales, $\mu/\dot{\mu}$, inferred from the measured braking indices.
We find that the observational data is consistent with the theoretical prediction though 
the small number of data precludes a strong conclusion. Measurement of the transverse velocities of pulsars
B1509$-$58, J1846$-$0258, J1119$-$6127 and J1734$-$3333 would increase the number of the data and strongly contribute to understanding
whether pulsar fields grow following fallback accretion.
\end{abstract}



\section{INTRODUCTION}

Soon after the discovery of radio pulsars \citep{hew68} their nature as rapidly 
rotating highly magnetized neutron stars (NSs) radiating at the expense of their 
rotational energy was established \citep{gol68}.
A dimensionless parameter related to the spin-down torques on these 
rotationally powered pulsars (RPPs) is the braking index defined  operationally as 
$n \equiv \nu \ddot{\nu}/\dot{\nu}^2$, where $\nu $ is the spin 
frequency and dots represent time derivatives. 
The value of $n$ should be 3 if RPPs are spinning down with magnetic dipole 
radiation $\dot{\nu}\propto -\mu^2 \nu^3$
where $\mu$ is the magnetic dipole moment. 
Most of the measured pulsar braking indices \citep{lyn93,liv07,wel11,roy12} are close to 3 
but slightly less as shown in Table~1 which indicate that another process also contributes
to the MDR in braking these objects. The recently measured braking index of PSR J1734$-$3333 
as $n=0.9 \pm 0.2$ \citep{esp11} and that of J0537$-$6910 as $n=-1.5$ \citep{mid06} together 
with the earlier measurement as $n=1.4\pm 0.2$ of the Vela pulsar \citep{lyn96} imply that 
this process might severely alter the spin history of young neutron stars.

\begin{table*}
\begin{minipage}{160mm}
\caption{Pulsars with accurately measured braking indices.
	\label{ta:brakings}}
\begin{tabular}{lcccccl}
\hline
Pulsar            & $\nu$     & $\tau_{c}$      & $n$         & $\tau_{\mu}$  & $v_{\perp}$      & References \\
                  & (Hz)      & (kyr)           &           & (kyr)         & (km s$^{-1}$)    &        \\
\hline 
B0531$+$21(Crab)  & 30.225    & 1.2399          & 2.51(1)   & 10.1(2)       & 140$\pm$8        & \cite{lyn93,ng06}   \\
B0833$-$45(Vela)  & 11.2(5)   & 11.303          & 1.4(2)    & 28(4)         & 62$\pm$2         & \cite{lyn96,ng07} \\
J1833$-$1034      & 16.159    & 4.8535          & 1.8569(6) & 16.984(9)     & 125$\pm$30       & \cite{roy12,ng07}    \\
B0540$-$69        & 19.738    & 1.6763          & 2.087(7)  & 7.34(6)       & 1300$\pm$612     & \cite{gra11,ng07} \\
J0537$-$6910      & 62.038    & 4.9743          & -1.5      & 4.422         & 634$\pm$50       & \cite{mid06,ng07}     \\
B1509$-$58        & 6.6115    & 1.5648          & 2.832(3)  & 37.3(7)       & $*$              & \cite{liv11_1509}     \\
J1846$-$0258      & 3.0743    & 0.72736         & 2.65(1)   & 8.3(2)        & $*$              & \cite{liv07}    \\
J1119$-$6127      & 2.4512    & 1.6078          & 2.684(2)  & 20.4(1)       & $*$              & \cite{wel11}    \\
J1734$-$3333      & 0.85518   & 81.280          & 0.9(2)    & 0.16(2)       & $*$              & \cite{esp11}    \\
\hline
\end{tabular}
\medskip  \\
Numbers in parenthesis are last digit errors.
\end{minipage}
\end{table*}

Different mechanisms have been invoked \citep[e.g.][]{mel97,men01,wu03,ho12} for addressing what makes 
the braking indices of RPPs less than 3. 
An early suggestion \citep{bla88} which has been recently advocated \citep{ber11,esp11} 
is that the braking indices of RPPs are less than 3 because their magnetic fields are growing in time.
If the magnetic dipole moment of a pulsar is changing in time, the braking index becomes
\begin{equation}
n = 3 + 2\frac{\dot{\mu}}{\mu}\frac{\nu}{\dot{\nu}} \ \ ,
\label{n1}
\end{equation}
\citep[e.g.][]{cha89} where $\mu$ is the magnetic dipole moment of the NS. 
The characteristic time-scale for the growth of the magnetic dipole moment, $\tau_{\mu} \equiv \mu/\dot{\mu}$, is then
\begin{equation}
\tau_{\mu}=\tau_c \frac{4}{3 - n} \ \ ,
\label{n2}
\end{equation}
where  $\tau_c \equiv -\nu/2\dot{\nu}$ is the characteristic spin-down age.
We have assumed here that the field growth is the dominant process modifying the braking index from its value 3 and the effect of other processes, e.g.\ those of magnetospheric currents \citep{con06,spi06} is small. The field growth timescales inferred from Equation~(\ref{n2}) are shown in Table 1. The wide range of $\tau_{\mu}$ values, 
from hundreds to millions of years, indicate that explaining the observed values of braking indices by growing magnetic dipole fields requires a process which could accomodate such different timescales. 

The magnetic dipole moment of a RPP could be growing because it was submerged \citep{mus95,yau95,gep99} in the crust due to fallback accretion  \citep{col71,zel72,che89} following the supernova explosion. In this work we predict from this hypothesis that pulsars which get large kick velocities during their birth should accrete less amount of matter, have shallower field burial and so shorter field growth time-scales. 
We plot field growth time-scales
$\tau_{\mu}$ as inferred from the measured braking indices via Equation (\ref{n2}) against the measured transverse velocities. We derive a relation between the Ohmic timescale for field evolution in the crust and the mass of the accreted matter finally relating the latter to the space velocity of pulsars. The observational data, low in number, marginally support the theoretical prediction favoring the idea that the magnetic field of young pulsars could be growing as a consequence of field diffusion to the surface following fallback accretion.

\section{A RELATION BETWEEN SPACE VELOCITY AND FIELD GROWTH TIMESCALE}

A common ingredient of supernova models is the fallback of some material  that can not reach the escape velocity \citep{col71,zel72,che89}. 
According to \citet{che89} the fallback due to the reversed shock in SN87A reached to the 
surface of the neutron star 2 hours after bounce.
The initial accretion rate of fallback matter can be very large. For SN87A the initial accretion rate is estimated \citep{che89} to be $350\,M_{\odot}\mathrm{yr}^{-1}$. 
The amount of matter that reaches the surface and the time required for this to happen varies depending on many details like the initial density of the surrounding medium,
relative velocity of the NS with respect to the medium, the magnetic field and spin frequency of the NS before fallback \citep{col96}.

The rapid accretion of fallback matter can bury the preexisting magnetic field that was formed in the proto-NS stage \citep{mus95,yau95,mus96,gep99,ho11,vig12,ber12}. 
The field would then diffuse back to the surface in an Ohmic time-scale 
$\tau_{\rm Ohm}=(\Delta R)^2/\eta $ where $\Delta R \sim 0.1$ km is the depth of submergence and $\eta \equiv c^2/4\pi \sigma $ is the magnetic diffusivity. 
The ohmic time-scale varies within a large range, hundreds to millions of years, depending on the conductivity, $\sigma$ of the crust and $\Delta R$ which, in turn, depends on the amount of fallback, $\Delta M$.
If the initial accretion rate is very large the magnetic field is submerged rapidly, and the amount of accreted matter is determined by the density of the medium and the velocity of the NS.

Old NSs in low mass X-ray binaries have magnetic fields $\sim 10^9$ G which is three orders of magnitude smaller than fields inferred for young NSs in high mass X-ray binaries and young isolated radio pulsars. The magnetic moments of NSs in binary systems is inversely correlated with accretion history \citep{taa86} supporting the
recycling scenario \citep{bis74,bis76,alp82,rad82} which suggests that the millisecond radio pulsars are spun up in x-ray binaries. 
Three distinct mechanisms have been suggested for the reduction of the magnetic field: (i) accretion induced heating decreases the conductivity of the star thus leading to the accelerated ohmic decay \citep{kon97,gep94,urp97}; (ii) vortex-fluxoid interactions in the superconducting core \citep{mus85,sri90}; and (iii) magnetic screening or burial \citep{bis74,che98,kon97,pay04,wan12}.
A well known problem with the field burial scenario is that the buried magnetic field will be prone to instabilities that rapidly will overturn the field \citep{lit01,vig08,muk12}. This problem, a solution of which  is beyond the scope of this \emph{letter}, is also relevant for fields buried by fallback accretion. We simply assume that whatever mechanism suppresses the instabilities in the case of binary accretion could also work in the case 
of fallback accretion.

\begin{figure*}
 \includegraphics[width=1\textwidth]{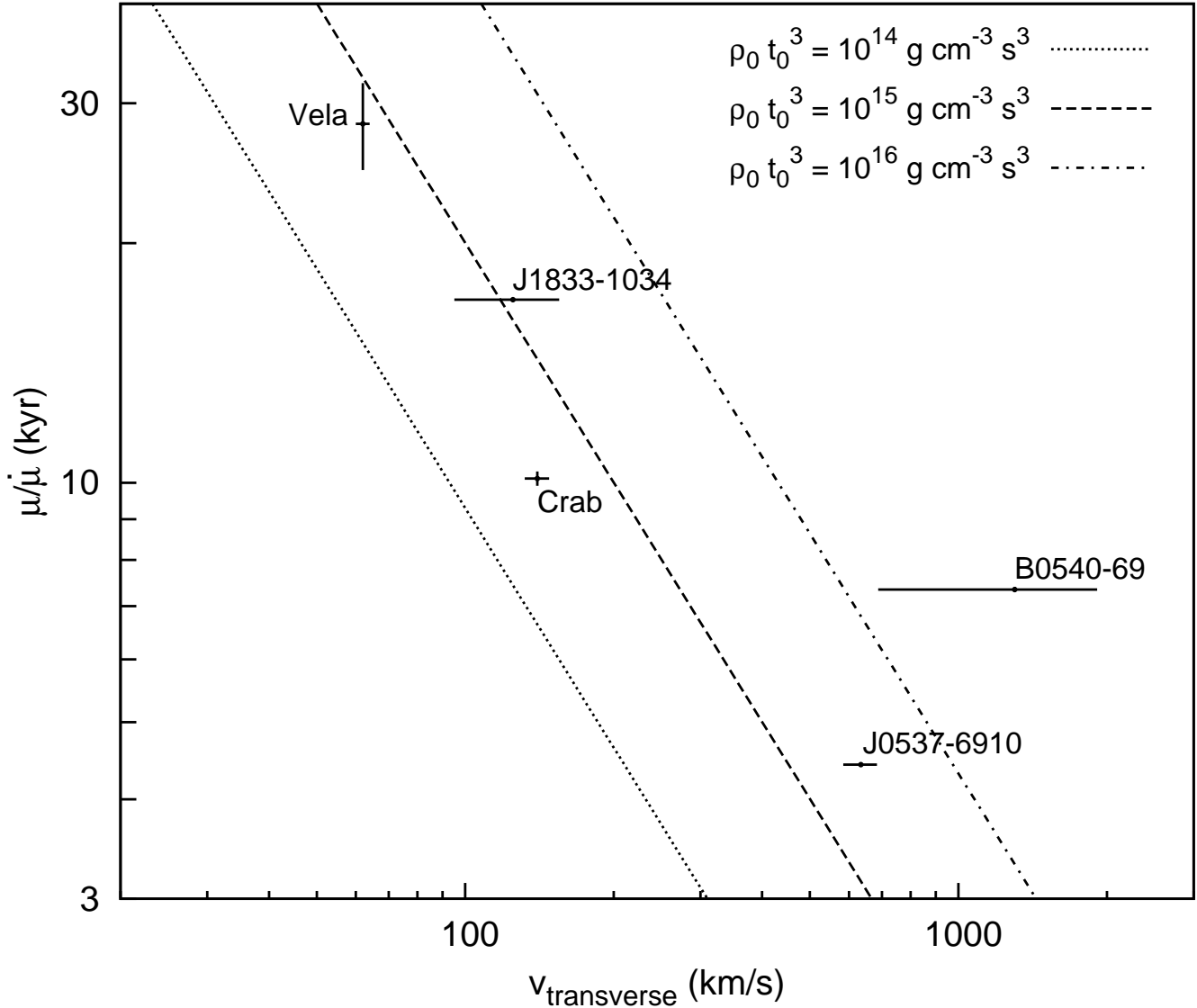}
 \caption{The relation between the measured transverse velocities $v_{\perp}$ and characteristic magnetic field growth time-scale $\tau_{\mu} \equiv \mu/\dot{\mu}$ inferred from measured braking indices via Equation (\ref{n2}) for the 5 pulsars, B0531$+$21(Crab), B0833$-$45(Vela), J1833$-$1034, B0540$-$69 and J0537$-$6910. These 5 objects form the subset of pulsars with accurately measured braking indices and measured transverse velocities. 
Also shown on the plot is the estimate of the Ohmic time-scale given in Eqn.(\ref{OHM}) depending on the space velocity $v$:
Dotted line stands for the case $\rho_0\ t_0^3 = {\rm 10^{14}\ gr\ cm^{-3}\ s^3}$,
dashed line stands for the case $\rho_0\ t_0^3 = {\rm  10^{15}\ gr\ cm^{-3}\ s^3}$ and
dashed-dotted line stands for the case $\rho_0\ t_0^3 = {\rm 10^{16}\ gr\ cm^{-3}\ s^3}$.
}
 \label{fig:1}                                                   
\end{figure*}

In binary accretion the burial of the magnetic field will proceed by the formation of a 
magnetically confined mountain \citep{woo82,ham83,muk12} of accreted plasma on the polar caps which then spreads laterally thus transporting the 
magnetic flux towards the equator \citep{che98}. The mountain then relaxes \citep{vig09} due to finite resistivity and sinks  while the accreted plasma becomes 
a part of the crust \citep{pay04,pay07,cho02,vig08,wet10}. Simply put, the burial of the field in a rather stable configuration requires the break down of spherical symmetry which is achieved by the modulation of the flow by the magnetosphere. As we show in the following, for the supercritical accretion rates expected to prevail at the initial stages of fallback accretion the magnetosphere, overwhelmed by the accretion flow, can not channel the inflowing matter to the polar caps. In this case the spherical symmetry is broken down by the asymmetry due to the space velocity of the pulsar i.e.\ the face of the NS in the direction of motion is subject to higher accretion rate than the opposite face. We expect in this case that the spin of the NS will convert the magnetic flux in the pre-existing dipole field to the toroidal component in the equator. To our knowledge burial of the magnetic field subject to Bondi-Hoyle accretion has not been explored numerically yet.

For a depth at which the density is greater than $10^6$ g cm$^{-3}$ the electrons in the crust are ultra-relativistic and the relativity parameter 
$x_{\rm F} \equiv p_{\rm F}/m_{\rm e} c \gg 1$ where $p_{\rm F}$ is the Fermi momentum and $m_{\rm e}$ is the electron mass. 
At the age of $\sim 10^4$ years which is the order of magnitude age of most of the considered pulsars with measured braking indices, $T \gtrsim 10^8$ K at the center of the 
star  and $T \gtrsim 10^6$ K at the surface. In such regimes the conductivity is determined by 
the electron-ion scattering in the melted layer at the surface and electron-phonon scattering 
in the solid crust 
\begin{equation}
\sigma_{\rm e-ph}=1.21\times 10^{22} x_{\rm F}^2 T_6^{-1}
\end{equation}
\citep{yak80} where $T_6 \equiv T/10^6$ K and we assumed $x_{\rm F} \gg 1$ \citep{urp94}. At lower temperatures conductivity will depend on the impurity scattering \citep{flo76} which
is independent of temperature.

To a good approximation the mass $\Delta M$ and thickness $\Delta R$ of the accreted layer are small fractions of the total mass, $M$ and radius $R$ of the  star, respectively.
Within this approximation the relativity parameter depends 
on $z\equiv \Delta R/H_{\rm R}$ as $x_{\rm F}=\sqrt{z(z+2)}$  \citep{urp79}  where $H_{\rm R}=m_{\rm e} c^2/g \mu_{\rm e} m_{\rm p}$ is the scale-height. 
Here $g=GMR^{-2}(1-2GM/Rc^2)^{-1}$, $\mu_{\rm e}=A/Z$ ($A$ is the atomic weight) and $m_{\rm p}$ is the proton mass. For $M=1.4M_{\odot}$, $R=10$ km and $\mu_{\rm e}=2$ one obtains $H_{\rm R}=7.25$ m.
The mass of the crust region of thickness $\Delta R$ is $\Delta M \simeq 4\pi R^{2}\int_{R-\Delta R}^{R}\rho (r)dr$ where $\rho$ is the density. 
By using the hydrostatic equilibrium equation \citep{opp39} this can be written as $\Delta M=(4\pi R^2/g)P_{\rm B}$ where $P_{\rm B}$ is the pressure at the bottom of the layer. For ultra-relativistic electrons $P=P_0 x_{\rm F}^4$ where $P_0=(2\pi/3)m_{\rm e}c^2 (m_{\rm e}c/h)^3=1.2\times 10^{23}$ dyne cm$^{-2}$. 
If one approximates $\sqrt{z(z+2)} \sim z$  as appropriate for $x_{\rm F} \gg 1$, then 
\begin{equation}
\Delta R = H_{\rm R}\left(\frac{g \Delta M}{4\pi R^2 P_0}\right)^{1/4} \simeq 77.5\, {\rm m}\, \left(\frac{\Delta M}{10^{-6}M_{\odot}}\right)^{1/4}.
\end{equation}
relates the accreted mass to the thickness of the resulting layer.
Assuming the conductivity due to electron-phonon scattering dominates we find the Ohmic time scale to be
\begin{equation}
\tau_{\rm Ohm} =36.7\, {\rm kyr}\, \left(\frac{\Delta M}{10^{-6}M_{\odot}}\right) T_6^{-1}.
\end{equation}

As the mass of the NS increases by accretion its radius becomes smaller and hence $g$ increases. 
The relation $\Delta M=(4\pi R^2/g)P_{\rm B}$ implies that the mass of the crust will 
decrease meaning that the difference is to be assimilated to the core \citep{kon97}. 
The movement of the current-carrying parts to larger depths where conductivity is larger will increase 
the Ohmic timescale and finally will lead to the freezing of the field if superconducting core is reached. 
We ignore this effect here as we infer that the amount of fallback mass accreted by any of the objects 
considered in this work is sufficiently small.
Yet we think the mechanism could be important for central compact objects like Cas A
which appear to have very small magnetic fields and could have relevance to 
the hidden magnetic field scenario \citep{vig12,ho11,sha12}.

The newborn neutron star will be accreting from a uniformly expanding medium where density 
decreases as $\rho \propto t^{-3}$ \citep{che89}.  The neutron star moving with velocity $v$ in this medium is subject to a mass inflow rate of $\dot{M}_{\rm in}=4\pi (GM)^2 \rho v^{-3}$ \citep{hoy39}. We combine these together as
\begin{equation}
\dot{M}_{\rm in} =\dot{M}_0 \left(1 + \frac{t}{t_0} \right)^{-3}, \qquad \dot{M}_0 = \frac{4\pi G^2 M^2 \rho_0}{v^3}
\label{mdot_in}
\end{equation}
where $\rho_0$ is the initial density of the medium and $t_0$ is the dynamical time-scale at which the accretion rate declines. We use the symmetry of the fluid dynamic equations under translations in time to get rid of the singularity at $t=0$ \citep{ert09}. Initially, the mass inflow rate can be very high and the amount of matter accreting onto the neutron star will be Eddington limited ($\dot{M}_{\ast}=\dot{M}_{\rm E}$) until
$t_1 = t_0 [ (\dot{M}_0/\dot{M}_{\rm E})^{1/3} -1  ]$
at which $\dot{M}_{\rm in}$ drops down to $\dot{M}_{\rm E}\equiv L_{\rm E}R/GM$ where $L_{\rm E}=4\pi GMm_{\rm p} c/\sigma_{\rm T}$ is the Eddington luminosity.

The magnetospheric radius is determined by the Alfv\'{e}n radius%
\begin{equation}
R_{\mathrm{A}}=\left( \frac{\mu ^{2}}{\sqrt{2GM}\dot{M}_{\mathrm{in}}}%
\right) ^{2/7}
\end{equation}
though this is very likely modified by the radiation pressure due to super-Eddington mass infall rates at the initial stage.
Initial value of the Alfv\'{e}n radius, referring Equation~(\ref{mdot_in}), is%
\begin{equation}
R_{\mathrm{A0}}=1\times 10^{6}\, {\rm cm}\,\, \left(\frac{\rho _{0}}{10^{-7}\, {\rm g\, cm^{-3}}}\right )^{-2/7}\mu
_{30}^{4/7}v_{100}^{6/7}
\end{equation}%
where $v_{100}\equiv v/(\rm 100\ km\ s^{-1})$ and we assumed $M=1.4 M_{\odot}$. 
For $\rho_0 > 10^{-7}\, {\rm g\, cm^{-3}}$, the initial value of the Alfv\'{e}n radius is smaller than the radius 
of the star suggesting that the magnetospheric radius is dynamically not important at the initial rapid accretion stage. In fact the initial accretion rate of fallback may be hypercritical i.e.\ photons are trapped in the accretion flow and the energy released in the accretion process is lost by neutrino emission \citep[see e.g.\ ][]{che89,ber11,ber12} as signaled by the divergence of $\dot{M}_{\rm in}$ for $v=0$. We do not assume this to be the case for the pulsars with large initial kicks that we consider in this work.
Even with the more modest initial accretion rates we consider, the NS starts
its life in the accretion mode and the propeller mechanism does not work at the initial stage at least not before most of the fallback matter accretes. Relying on this we assume that the spin and magnetic field of the NS will not change the amount of accreted matter significantly unless they have magnetar-like initial values, $\mu \sim 10^{33}$ G and $P_0 \sim 1$ ms.

The accretion rate onto the neutron star, for $t>t_1 \gg t_0$  is $\dot{M}_{\ast}=\dot{M}_{\rm E}(t/t_1)^{-3}$. This implies that the total mass accreted onto the neutron star will be $\Delta M = \frac32 \dot{M}_{\rm E}t_1$. Assuming $\dot{M}_0 \gg \dot{M}_{\rm E}$ we find $\Delta M = \frac32 \dot{M}_{\rm E}t_0  ( \dot{M}_0 / \dot{M}_{\rm E} )^{1/3}\propto v^{-1}$ and this leads to
\begin{equation}
\tau_{\rm Ohm} = 202\, {\rm kyr}  \left(\frac{\rho_0 t_0^3}{\rm 10^{15}\ g \ cm^{-3}\ s^3}\right)^{1/3} T_6^{-1} v_{100}^{-1}
\label{OHM}
\end{equation}
where we scale the poorly known quantities $\rho_0$ and $t_0$ together. As the fallback is expected to occur a few hours after the supernova explosion \citep{che89,ber12} typically $t_0 \sim 10^4$ s which means $\rho_0 \sim 10^3$ g cm$^{-3}$
would provide the observed time-scale for the field evolution.

As a test of this relation $\tau_{\rm Ohm} \propto v^{-1}$ we check the relation between $\tau_{\mu}$ inferred from braking indices via Equation (\ref{n2}) and measured transverse velocities $v_{\perp}$ expecting $\tau_{\mu} \sim \tau_{\rm Ohm}$ and $v_{\perp} \sim v$. Only 9 RPPs have accurately measured braking indices that would allow for inferring the time-scale $\tau_{\mu}$ for the growth of their magnetic moment 
via Equation~(\ref{n2}). Of these RPPs, there are only 5 with measured transverse velocities as shown in Table~1. 
In Figure~\ref{fig:1} we plot $\tau_{\mu}$  versus $v_{\perp}$ of these objects together with the estimate of the Ohmic time-scale in Eqn.(\ref{OHM}) depending on the space velocity $v$. 
Although the number of data is small, an inverse relation between the field growth time-scale and the transverse velocities can still be noticed as a result of the strong dependence of accretion rate on $v$. The scattering in the data is most likely the result of  differences in the temperatures of the pulsars, the initial density $\rho_0$ of the surrounding media, and possibly the weaker factors like the initial spin and magnetic field of the neutron star. Note also that $\tau_{\mu}$ does not necessarily coincide with $\tau_{\rm Ohm}$ if the field growth is not exponential. Add to this that $v_{\perp}$ may not be a good estimate of the space velocity $v$ if the radial component of the velocity is not negligible compared to $v_{\perp}$. Given these deteriorating
factors the relation obtained with 5 data can be considered promising.

\section{DISCUSSION} 

As a prediction of the fallback submergence and post-fallback growth of magnetic fields, we have derived the relation 
$\tau_{\rm Ohm} \propto v^{-1}$ between the space velocity $v$ and Ohmic time-scale for the diffusion of a buried magnetic field, $\tau_{\rm Ohm}$.  We looked for a similar relation between the measured transverse velocities
$v_{\perp}$ and characteristic field growth time-scales $\tau_{\mu} \equiv \mu/\dot{\mu}$ inferred from the measured braking indices. Such an inverse relation between the kick velocity and the field-growth time scale of pulsars is a prediction of only post-fallback magnetic field growth model and not of any other model.

As the number of pulsars with both measured braking indices and transverse velocities is small (only 5), it is not possible to claim for a strong relation.  Yet we  found the result promising given there are many factors that 
could deteriorate the predicted simple relation.
It appears from this analysis that measuring the transverse velocities of pulsars 
B1509$-$58, J1846$-$0258, J1119$-$6127 and J1734$-$3333 would almost double the number of data and allow for a stronger conclusions for the model.

Our estimations indicate that the depth of field submergence for the considered population of young neutron stars is very shallow, possibly leading to the selection of these objects as rotationally powered pulsars. Smaller space velocities result with larger amount of accretion
and very long diffusion time-scales of buried magnetic fields leading to young neutron stars not appearing as rotationally powered pulsars. This hidden magnetic field scenario is recently studied by \citet{vig12,ho11,sha12}. This, as well as a more accurate numerical analysis of the field growth of pulsars, is the subject of a following work.

\section*{Acknowledgments}
K.\ Y.\ E.\ acknowledges support from the Faculty of Science and Letters of \.{I}stanbul Technical University.
K.\ Y.\ E.\ thanks A.\ Patruno for hospitality during his visit to Amsterdam and M.\ A.\ Alpar for his support 
and careful reading of an early version of the manuscript.
We thank T.\ Tauris and W.\ Ho for helpful comments. We thank the anonymous referee for comments that helped to improve the paper.

\bibliographystyle{mn2e}

\label{lastpage}

\end{document}